\documentclass[onecolumn,showkeys,notitlepage]{revtex4-1} 

\usepackage{braket}
\usepackage{graphicx}
\usepackage{amsfonts}

\usepackage{etoolbox}

\begin{document}

\title{CUBIT: Capacitive qUantum BIT}

\author{Sina Khorasani} 
\affiliation{Vienna Center for Quantum Science and Technology, University of Vienna, 1090 Vienna, Austria; sina.khorasani@ieee.org}

\date{June 17, 2018}

\begin{abstract}
In this letter, it is proposed that cryogenic quantum bits can operate based on the nonlinearity due to the quantum capacitance of two-dimensional Dirac materials, and in particular graphene. The anharmonicity of a typical superconducting quantum bit is calculated, and the sensitivity of quantum bit frequency and anharmonicity with respect to temperature are found. Reasonable estimates reveal that a careful fabrication process can reveal   expected properties,   putting the context of quantum computing hardware into new perspectives.
\end{abstract}


\maketitle

\section{Introduction}

We discuss a new type of qubit with integrability, based on a nonlinear quantum capacitor (QC) made of   two-dimensional materials~\cite{Ref1}, graphene (Gr), and boron nitride (BN) and their heterostructures~\cite{Ref1a,Ref1b}. Without exception, all existing superconducting qubits~\cite{Ref2,Ref7,Ref3,Ref7a,Ref7b,Ref7c} rely on the nonlinear inductance of Josephson junctions (JJs). Here, the~nonlinear inductance is replaced by the nonlinear capacitance coming from two Gr monolayers separated by a multilayer BN, which acts as a potential barrier and forms a parallel plate capacitance~\cite{Ref1}. 

The nonlinearity of such a quantum capacitor stems from the two-dimensionality of crystal combined with the Dirac cone in its band structure, where a square-root correspondence between the Fermi energy $E_F$ and external voltage $V$ is established. There is no fundamental reason why other 2D Dirac materials, such as silicene and germanene~\cite{Si,Ge,Dirac,Small} could not be used for this purpose. 

On the practical side, two major obstacles seem to influence the design. One is the compatibility of Gr with superconducting circuits, and the next one is the potential puddles arising from crystal imperfections and impurities. Recent progress on large-scale and near-perfect crystalline growth of Gr~\cite{Smirnov}, which is equally applicable to BN, should in principle resolve the issues tied to the puddles to a great extent. Additionally, recent discoveries on the existence of  the p-wave-induced phase of superconductivity in Gr~\cite{Ferrari} as well as  the unconventional superconducting phase of  the twisted bilayer Gr at the magic angle of $19.19\text{mRad}$~\cite{Herrero} raises hopes for  the practicality of such a design. The latter  offers a large decrease in Fermi velocity at the vicinity of Dirac cone, which helps in a significant reduction in qubit size as well as  the relaxation of stringing limits imposed by the fabrication technology. 

It is the purpose of this letter to establish the feasibility of \textit{CUBIT}, which we use to refer to the nonlinear Capacitive qUantum BIT. While we present the numerical design of a preliminary CUBIT, the~fundamental aspects and practicality issues   involved are also discussed. Another potentially useful application of nonlinear quantum capacitance could be in the cryogenic quantum-limited parametric amplifiers. 

\section{Results}
\begin{figure}[ht!]
	\centering
	\includegraphics[width=4in]{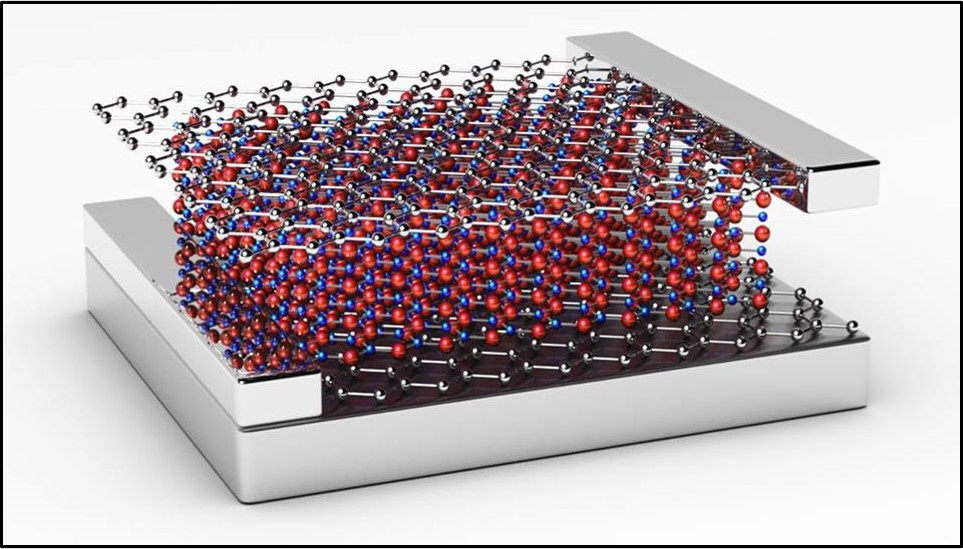}
	\caption{Graphene/boron nitride/graphene sandwich structure. Wide aluminum contacts are attached across the edges. \label{Fig1}}
\end{figure}

This sandwich structure shown in Figure \ref{Fig1} is routinely used owing to its peculiar electronic and optical properties~\cite{Ref4,Ref5}. Quite interestingly, its application to a quantum-dot charge qubit design has been also demonstrated~\cite{Novoselov}, which is however incompatible with the conventional superconducting quantum circuits.

For the BN thickness of 3  to 70~nm, the~contribution of the  geometric parallel plate capacitance of the structure  can be neglected, and the quantum capacitance is given by~\cite{Ref2,Ref2a,Ref2b}
\begin{equation}
\label{eq1}
C_Q=\frac{2e^2 S k_B T}{\pi(\hbar v_F)^2 } \ln\left[2\left(1+\cosh\frac{E_F}{k_B T}\right)\right] 
\end{equation}
where $v_F$ is the Fermi velocity of graphene, $k_B$ is the Boltzmann’s constant, $e$ is the electronic charge, $S$~is the total geometric 2D area of capacitor, and $T$ is the absolute temperature. Evidently, the~quantum capacitance is nonlinear since it depends on the applied voltage $V=2E_F/e$. 

The recently discovered superconductivity in  the  twisted bilayer Gr at the magic angle~\cite{Herrero}  offers a roughly 7-fold reduction in $v_F$. This can cause a significant reduction of the necessary capacitance area up to a factor of roughly $50$, since $C_Q\propto S/ v_F^2$, while maintaining the same level of anharmonicity $A$. However, such a structure needs voltage bias and   cannot accommodate a large number of carriers. That would necessitate a serious look and a separate study.

\begin{figure}[ht!]
	\centering
	\includegraphics[width=2.5in]{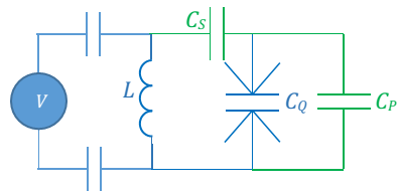}
	\caption{Illustration of the basic capacitive quantum bit design. The addition of series and parallel extra capacitors adds up to the design flexibility. \label{Fig2}}
\end{figure}

The bulk of the proposed qubit is a nonlinear anharmonic oscillator, which forbids transitioning from $\ket{0}$ to $\ket{n}; n>1$ states with a given fixed pump. This has been schematically shown in Figure \ref{Fig2}. Signal can be fed into the oscillator by inductive coupling, or capacitive coupling similar to the qubits based on Josephson junctions~\cite{Ref7,Ref6}. The strength of anharmonicity could be adjusted by temperature as well as other available design parameters of the layered capacitor. 

The total linear and nonlinear part of the QC can   then be expanded as a power series in terms of the applied voltage. By placing a linear inductor 
  across the QC, the~total Hamiltonian up to  the fifth order is found, with remarkably large anharmonicity $A$. It is possible to tweak the design flexibility by placing a shunt linear capacitor across the QC. It is found that $A$ rapidly increases as the temperature decreases and   is a strong function of the capacitance area~\cite{Ref1}, and temperatures of a few 0.1 K easily yield strong nonlinearity. Signal can be fed into the qubit oscillator by inductive or capacitive coupling, similar to the qubits based on JJs~\cite{Ref2}. 

It is possible to tweak the design flexibility by placing the linear capacitors $C_S$ and $C_P$, respectively, in series and across the quantum capacitor as shown in Figure \ref{Fig2} in green. This placement of the shunt capacitor $C_P$ also could relax the restriction on the maximum operation temperature given as $\hbar \omega>2k_B T$. It should be noted that the existence of extra capacitors decreases the large anharmonicity delivered by the quantum capacitance $C_Q$.

\begin{figure}[ht!]
	\centering
	\includegraphics[width=3.5in]{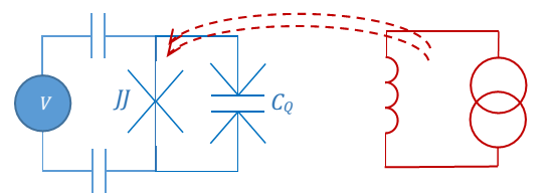}
	\caption{Illustration of a combined nonlinear quantum capacitance and Josephson junction (JJ) design. \label{Fig3}}
\end{figure}

Shown in Figure \ref{Fig3}, it is furthermore possible to combine the QC with JJs, where $A$ is caused by the combined effects of JJ and QC. Quite clearly, the~resonance frequency of the circuit is now also dependent on  the linear inductance of the JJ as well. A proper design can cause complete cancellation of or enhancement in the anharmonicity of $A$, leaving nonlinear terms of sixth and higher orders. Furthermore, positive or negative $A$ become now both accessible, using appropriate biasing of the JJs. 

The combined effects of such an anharmonic oscillator can be expressed by the Hamiltonian
\begin{equation}
\label{eq2}
\mathbb{H}=\hbar\omega\left(\hat{a}^\dagger\hat{a}+\frac{1}{2}\right)+\frac{1}{4}\hbar\omega(\alpha-\omega\tau) \left(\hat{a}^\dagger+\hat{a}\right)^4
\end{equation}
where $\alpha$ is the contribution of JJs, and $\tau$ is the characteristic time of nonlinear interaction for the quantum capacitor~\cite{Ref2}. This value has   already been estimated theoretically~\cite{Ref2} for a weakly nonlinear regime, which is not obviously a matter of interest in qubit design. The reason is that the anharmonicity should   actually be large to guarantee two-level operation of the oscillator, strictly allowing no more than 1 photon at the qubit frequency $\omega$ to survive in the oscillator. Therefore, in this study, $\tau$ has been numerically estimated by an accurate solution of the nonlinear Hamiltonian potential  and by 
 obtaining the energy eigenvalue levels. Energy eigenvalues were found using a recent semi-analytical method reported elsewhere~\cite{Basis}, which is at least as reliable as WKB but much easier to implement.

While it may seem at first that the nonlinearity is too large to be acceptable, it should be kept in mind that the quantum capacitance of an ideal Gr sheet tends to be singular with infinite derivative at zero temperature. At finite temperature, this singularity is less sharpened and less pronounced, which ultimately results in a large anharmonicity, inversely increasing with the third power of absolute temperature. When the Gr sheet is non-ideal because of potential puddles and other crystal imperfections,   this sharpness can reduce, and beyond a certain density of puddles, the~useful and available single-photon anharmonicity vanishes. This is   discussed   in Section \ref{Practical}.

Since the limiting cases and truncated expansions may not be sufficiently accurate,   the anharmonicity was studied with a completely numerical approach, where limiting expressions of the type reported in the previous study~\cite{Ref2} no longer hold true and significantly deviate from exact values. The numerical computations are rather elaborate; therefore, instead of  a description of the entire algorithm and minor details,  the Mathematica calculation packages have been made freely available to   readers as Supplementary Information.

As it goes with the analytical form of Hamiltonian,  analytical expansions such as Equation (\ref{eq2}) allow significant anharmonicity to appear when the rotating wave approximation is applied, that is, when~the time-independent terms are kept and the rest are dropped. This will still allow a significant anharmonic behavior to emerge in the form of an effective Kerr Hamiltonian, as long as $\tau\omega << 1$ can be violated. Yet, a Kerr Hamiltonian is accurate only up to the fourth order, while the nonlinear behavior of cubits implies that all even higher powers of number operators will appear. This will set restrictions on the accuracy and usefulness of Equation (\ref{eq2}), which ultimately necessitates a full numerical approach.

Since the structure has to   operate  inside a dilution refrigerator with some temperature instability and drift over time, and the fact that the quantum capacitance is a function of temperature, one should make an estimate of qubit frequency $\omega=2\pi f$ variations with temperature. 

\begin{figure}[ht!]
	\centering
	\includegraphics[width=3.5in]{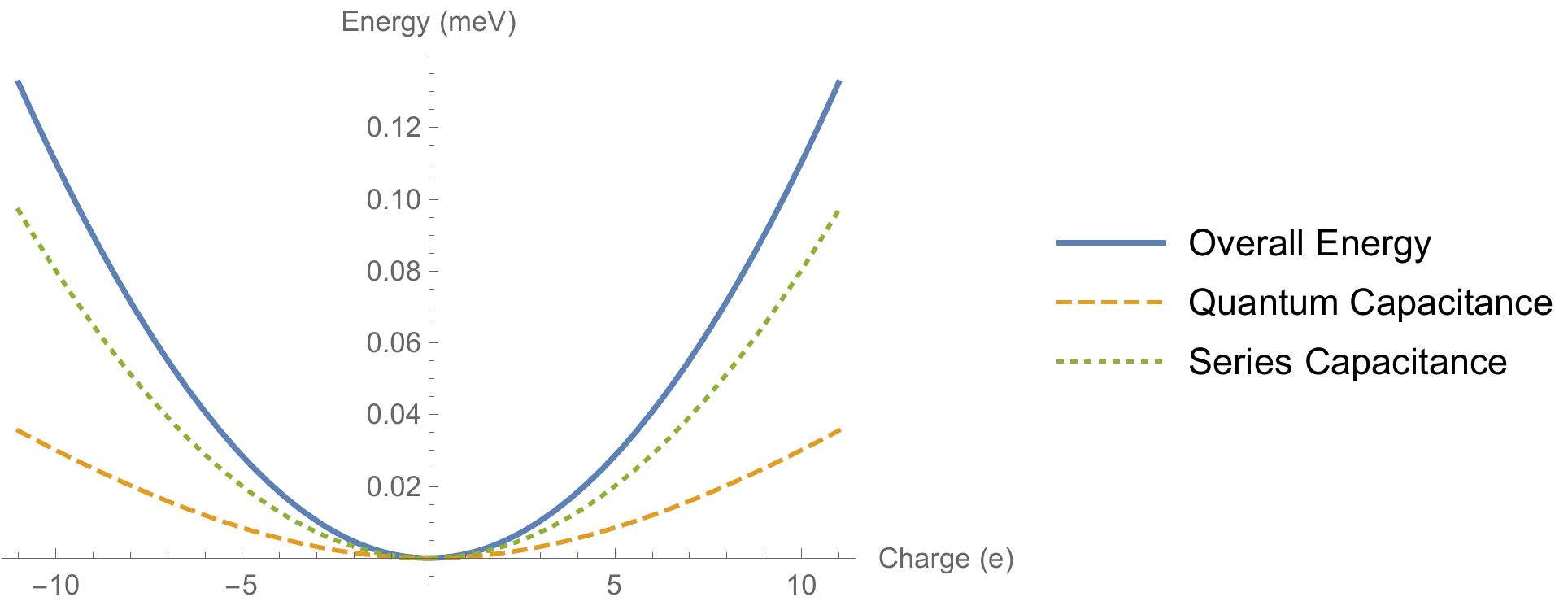}
	\includegraphics[width=2.2in]{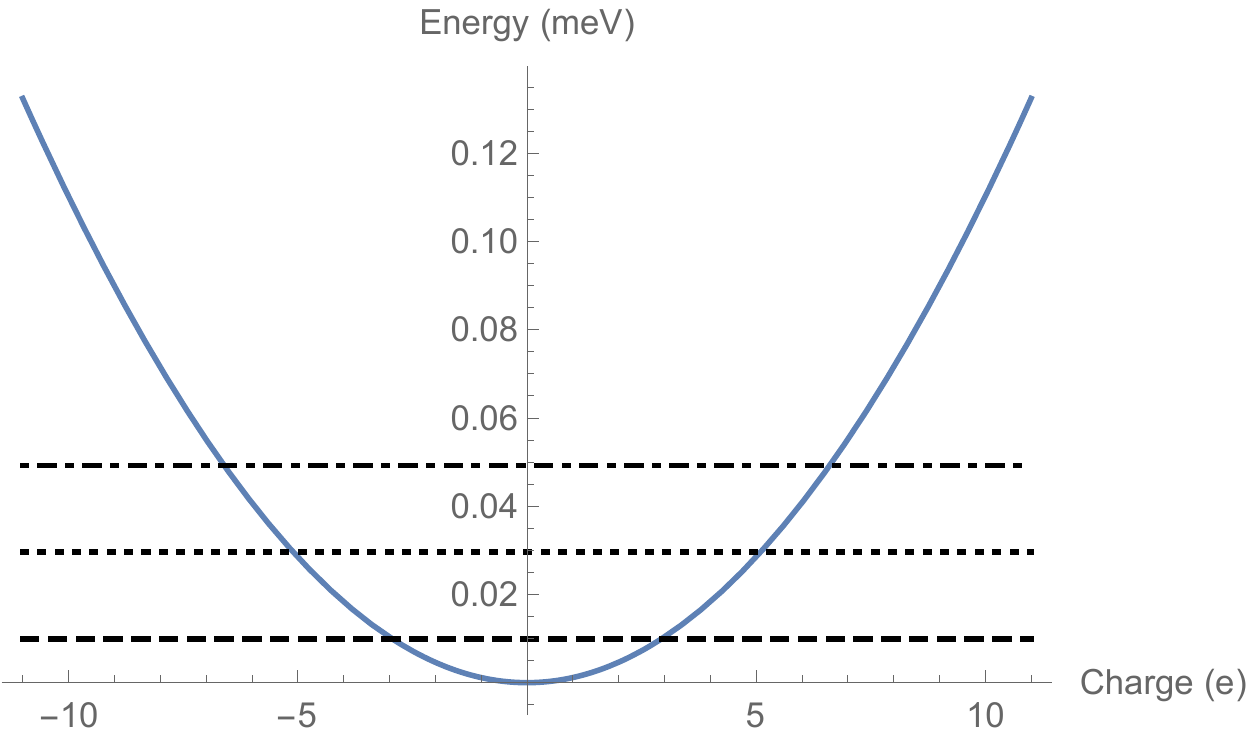}
	\caption{Calculation of energy levels of the anharmonic oscillator composed of nonlinear quantum capacitor and linear inductor shown in Figure \ref{Fig2}. \label{Fig4}}
\end{figure}

Additionally, the~proposed qubit as an anharmonic oscillator  is also dependent on the graphene area as well. One may put the nonlinear Hamiltonian of the whole qubit together and solve for the eigenstates, which correspondingly yield the energy eigenvalues and therefore the anharmonicity. A typical dependence of the potential energy of the anharmonic circuit, including contributions of the quantum and series capacitance looks like the following in Figure \ref{Fig4}.

\begin{figure}[ht!]
	\centering
	\includegraphics[width=2.9in]{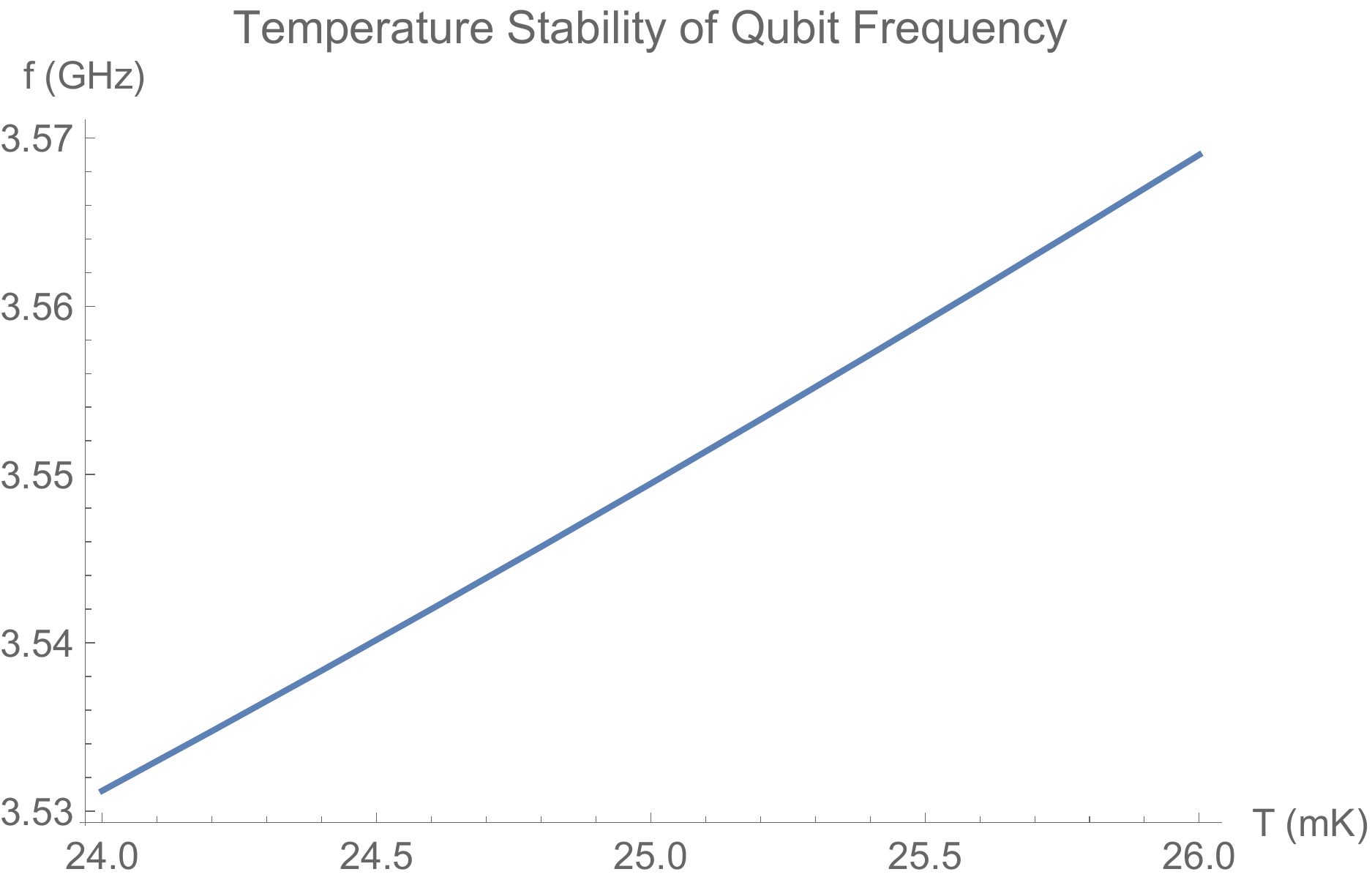}
	\includegraphics[width=2.9in]{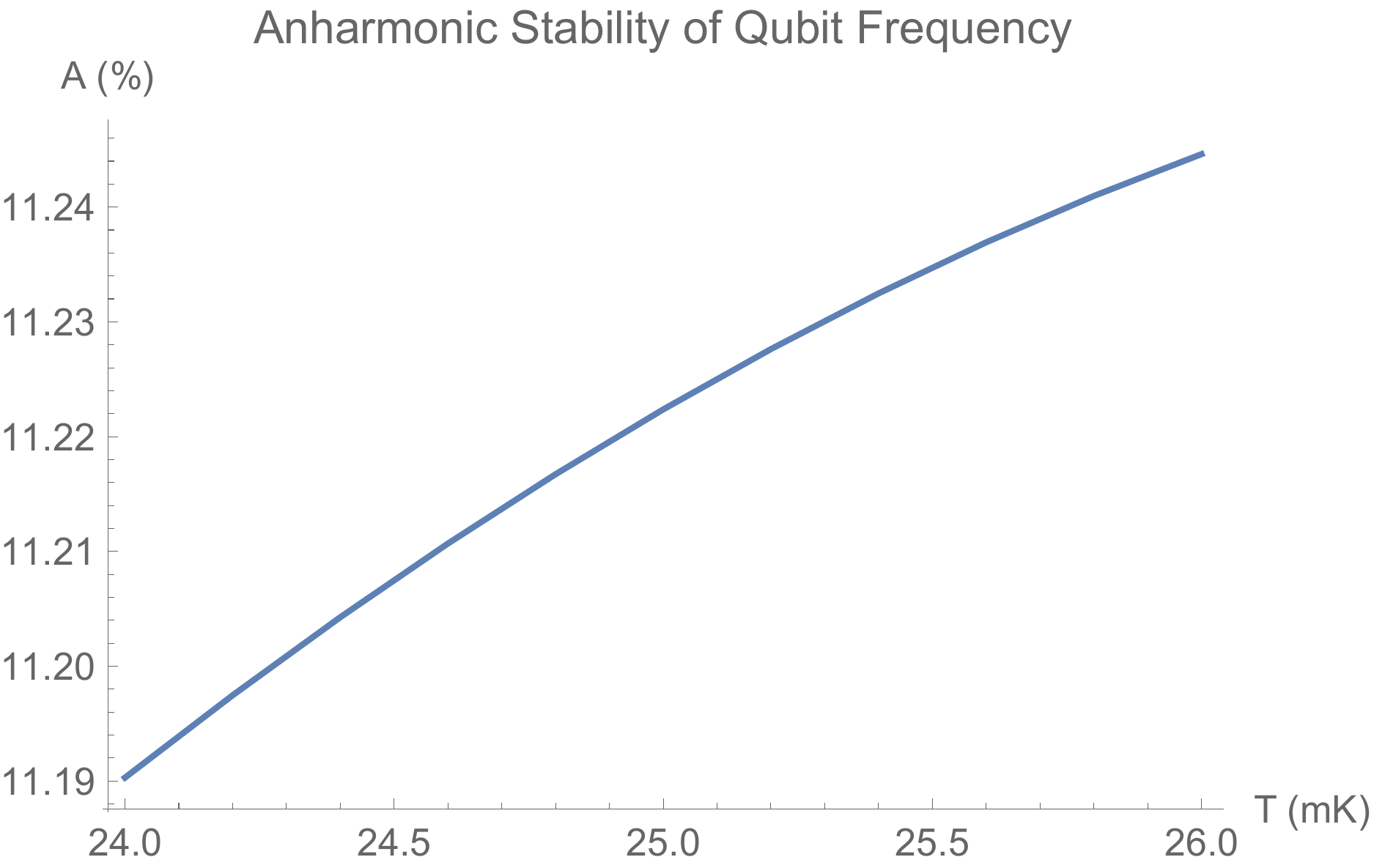}
	\caption{Variations of qubit frequency and anharmonicity versus operation temperature of the dilution fridge. \label{Fig5}}
\end{figure}

Since the oscillator is strongly anharmonic, the~ultimate actual transition frequency can be different from the initial value by design. This anharmonicity is not only a function of temperature but also a function of capacitor area, too. This dependence has been investigated as follows in Figure \ref{Fig5}.

\begin{figure}[ht!]
	\centering
	\includegraphics[width=2.9in]{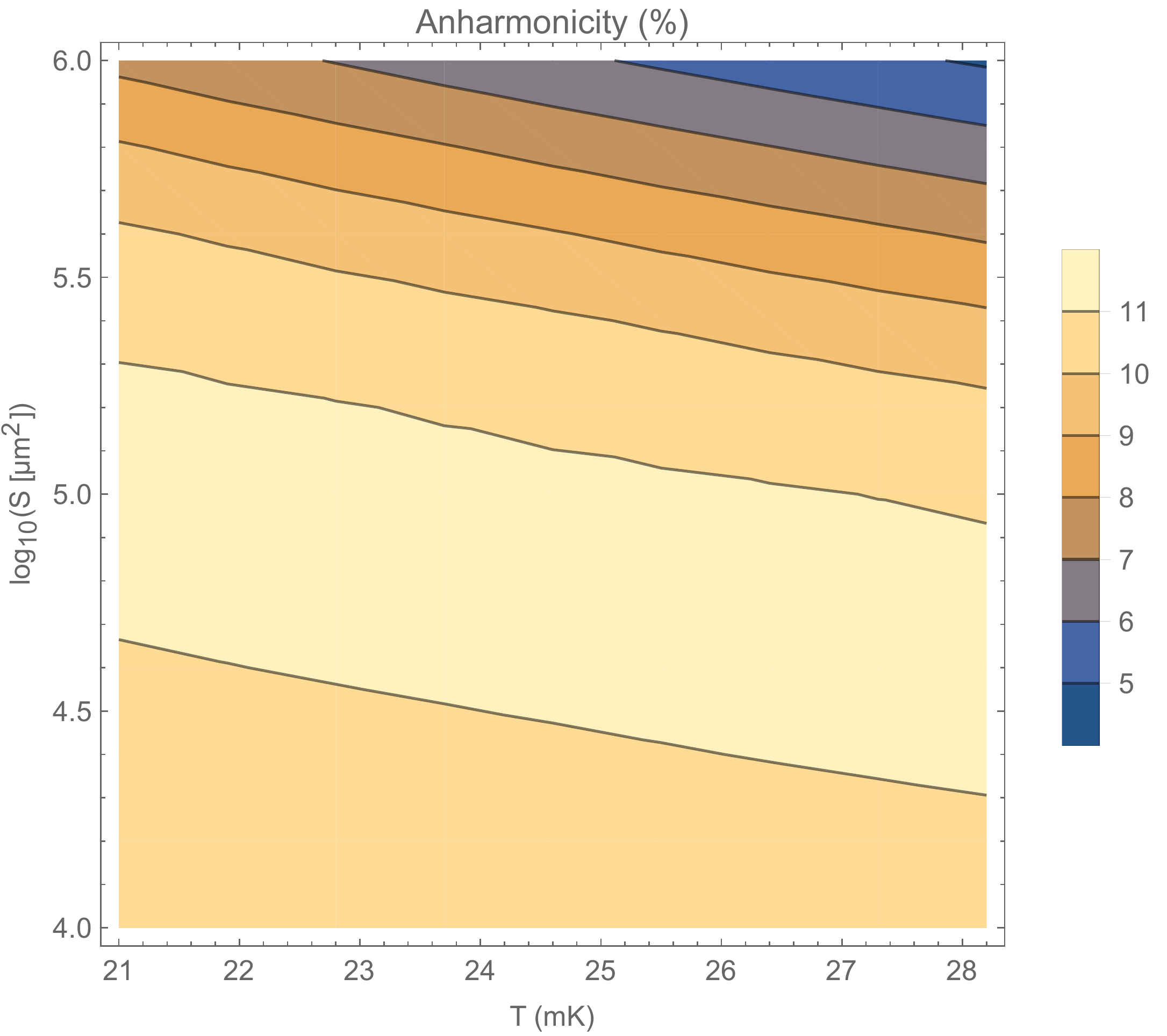}
	\includegraphics[width=2.9in]{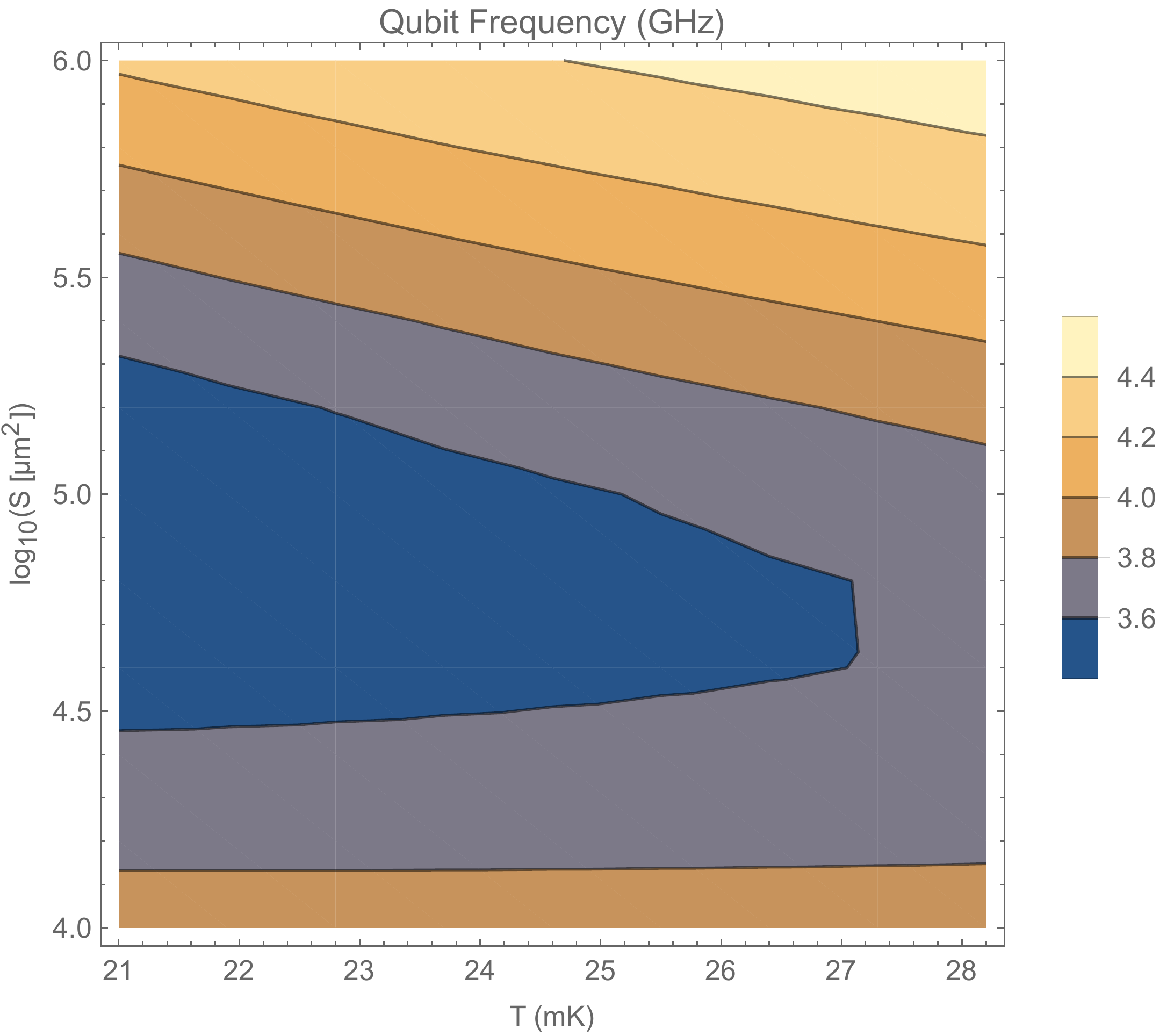}
	\caption{Contour plots of anharmonicity and qubit frequency versus temperature and qubit area. \label{Fig6}}
\end{figure}

The largest available range of anharmonicity occurs at the capacitor area of the order of \mbox{$5\times10^4$~$\mu\text{m}^2$} to $10^5$~$\mu\text{m}^2$. Hence, a desirable capacitor size could be chosen to be around 50 $\mu\text{m}\times 1~\text{mm}$, which together an inductor of 60 nH would yield a qubit frequency of $\omega=2\pi\times 3.55~\text{GHz}$. This area is evidently small enough to be considered for easy fabrication and   integration. The temperature sensitivity could be observed by inspection of the temperature-dependent variations of qubit frequency. This is illustrated in   Figure \ref{Fig6}.

This calculation gives the estimates of
\begin{eqnarray}
\frac{\partial f}{\partial T}&=19\frac{\text{MHz}}{\text{mK}}, \\ \nonumber
\frac{\partial A}{\partial T}&=0.027\frac{\%}{\text{mK}},
\end{eqnarray}
which at an operation temperature of 25~mK are equivalent to the operational sensitivities of
\begin{eqnarray}
S_f^T&=\frac{\partial f/f}{\partial T/T}\times 100\%=9.5\%  \\ \nonumber
S_A^T&=\frac{\partial A/A}{\partial T/T}\times 100\%=6.1\%.
\end{eqnarray}

Some design numbers are shown in Table \ref{Table1}, while redesigning with a series capacitor of 0.1  to 1~pF gives the data in Table \ref{Table2}.

\vspace{-12pt}

\begin{table}[ht!]
	\centering
	\caption{Qubit designs with no series capacitor $S=1~\text{mm}^2$.}
	\label{Table1}
	\begin{tabular}{ccccccc}
		\hline\hline
		\textbf{S (}\boldmath$\text{B}^2$\textbf{)} & \textbf{Design} \boldmath$\omega/2\pi$ \textbf{(GHz) }& \textbf{T (mK)}  & \boldmath$C_S$ \textbf{(fF)} & \textbf{Actual} \boldmath$\omega/2\pi$ \textbf{(GHz)}  & \textbf{A (}\boldmath$\%$\textbf{)}  \\ \hline
		1 &  2.5 & 25  & $-$ & 2.29 & 3.9   \\ 
		1 &  5  &  25 & $-$ & 4.41  & 6.04   \\ 
		1 &  10 &  25 & $-$ & 8.31  & 8.26 \\  \hline\hline
	\end{tabular}
\end{table}
\unskip
\begin{table}[ht!]
	\centering
	\caption{Qubit designs with a series capacitor $S=1~\text{mm}^2$.}
	\label{Table2}
	\begin{tabular}{ccccccc}
		\hline\hline
		\textbf{S (} \boldmath$\text{\textbf{mm}}^2$\textbf{)} & \textbf{Design}  \boldmath$\omega/2\pi$ \textbf{(GHz)} & \textbf{T (mK) } &  \boldmath$C_S$ \textbf{(fF)} & \textbf{Actual}  \boldmath$\omega/2\pi$ \textbf{(GHz)}  & \textbf{A (}\boldmath$\%$\textbf{)}  \\ \hline
		1 &  2.5 & 25  & 100 & 2.39 & 0.44   \\ 
		1 &  5  &  25 & 100 & 4.77  & 0.76   \\ 
		1 &  10 &  25 & 1000 & 8.71  & 5.67 \\  \hline\hline
	\end{tabular}
\end{table}

No surprise that the presence of the series capacitor should cause a reduction in anharmonicity $A$, since it makes the capacitive behavior of the whole circuit more linear. In general, increasing the frequency  while reducing the capacitor area seems to lead into an even more desirable set of numbers shown in Table \ref{Table3}.

\begin{table}[ht!]
	\centering
	\caption{Qubit designs with no series capacitor $S=0.1~\text{mm}^2$.}
	\label{Table3}
	\begin{tabular}{ccccccc}
		\hline\hline
		\textbf{S (}\boldmath$\text{\textbf{mm}}^2$\textbf{) }& \textbf{Design} \boldmath$\omega/2\pi$ \textbf{(GHz)} & \textbf{T (mK)}  & \boldmath$C_S$\textbf{ (fF)} & \textbf{Actual} \boldmath$\omega/2\pi$ \textbf{(GHz) } & \textbf{A (}\boldmath$\%$\textbf{) }\\ \hline
		0.1 &  10 & 25  & $-$ & 6.42 & 11.2   \\ 
		0.1 &  15  &  25 & $-$ & 11.1  & 9.02   \\ 
		0.1 &  20 &  25 & $-$ & 11.5  & 11.1 \\ \hline\hline
	\end{tabular}
\end{table}

Here, we ultimately choose the capacitor area of $S=5\times 10^4 \mu\text{m}^2$, which is $50 \mu\text{m}$ wide and 1~mm long, together with an inductor of $L=60 \text{nH}$, which yields the qubit frequency of $\omega=2\pi\times 3.55 \text{GHz}$, and anharmonicity of $A=11\%$. The zero-point voltage and carrier number at the qubit frequency is estimated to be around $V_{\rm zp}$ $=15\mu\text{V}$ and $n_{\rm zp}=1.7$. All contacts and connecting materials can be chosen to be aluminum, since not only does it easily superconduct  at the temperatures of interest  but is also a quite typical metal of choice in superconducting qubits.

It has to be mentioned again that employing the superconducting bilayer graphene~\cite{Herrero} can in principle cause up to a 50-fold reduction in qubit area. That would limit the typical qubit area only to $S=10^3~\mu\text{m}^2$, which is now a highly reasonable value.

\section{Practical Considerations}\label{Practical}
\vspace{-6pt}

\subsection{Potential Puddles}

In graphene, non-ideal impurities and defects at the interface do exist. Although their density can be minimized by exfoliation under high-vacuum and   encapsulations with 2D hexagonal BN, they~persist to certain densities. Traps form shallow potential kinks, named as puddles. 

Puddles are expected to be frozen and unmoving at ultralow temperatures. But their density and potential depth can be a matter of concern. For the quantum capacitance of Gr to survive, it is estimated that the puddle surface density must be under $N_{\rm pd} <10^8~\text{cm}^{-2}$, which is roughly 1 defect per every $10^8$ Carbon atom in Gr lattice. Furthermore, puddle potential energy depth must be bounded by $U_{\rm pd}<10~\text{meV}$. Values exceeding these bounds are undesirable and result in destroying the quantum capacitance property of Gr.

Typical values that are achieved in the experiments are not better than $10^{10}~\text{cm}^{-2}$ yet. While~this might seem a bit disappointing, the~recent remarkable progress in very large-scale crystalline CVD growth of Gr and 2D materials~\cite{Smirnov}  highlights the likelihood of this possibility, and even much better values to be attained soon. 

The superconducting twisted bilayer graphene at the magic angle~\cite{Herrero} can not only   result in a significant reduction of qubit size but also may largely relax the constraints on the puddle density. This~can be studied in depth either by ab initio approaches such as the density function theory (DFT) or~through a series of carefully conducted experiments.

\subsection{Zero-Point Fluctuations}

With regard to the effect of impurity charges, these appear as a background bias in the overall charge density. For a designed device to be practically useful at the single photon level, the~zero-point charge fluctuations should exceed the impurity charges. The correct way to obtain zero-point amplitudes is to look for the corresponding values that reproduce the energy of half-quanta $\frac{1}{2}\hbar \omega$.

\subsection{Decoherence and Dephasing}

Possible mechanisms for qubit decoherence and dephasing may be considered, which include but are not limited to spontaneous emission from excited states, enhancement of emission rate due to the Purcell enhancement (when the qubit is placed inside a high-Q microwave cavity), substrate dielectric losses, nonideal proximity effects of carrier transport in graphene, tunneling across the dielectric due to defects and periphery surface states, coupling to spurious modes and surface acoustic waves, charge noise, flux noise, and strain noise (which cause fluctuations and anisotropy in the Fermi velocity $v_F$). 

Some of these have already been studied for transmon qubits~\cite{Ref7,Ref6}, and their practical limits have been evaluated. Some have to be studied in great detail and require    deep theoretical study. A  $0-\pi$ qubit design strategy that offers inherent immunity with respect to decoherence has been introduced very recently \cite{0pi}. While this has yet to be tested experimentally, a topological dual of this strategy may in principle be applied to the anharmonic CUBIT under consideration here as well.

One has to keep in mind that superconducting qubits have already come a long way over the past 18 years, with their coherence times being initially only around 1~ns~\cite{Ref7}. Meanwhile, the coherence performance of transmons is now for all practical reasons being saturated to 0.1 ms for the best available designs and fabrication processes, such as the one being used in IBM-Q. That would already be   a $10^5$~fold improvement.

\subsection{Parametric Amplifiers}

An alternative potential use for this type of non-dissipative cryogenic nonlinear element could be in parametric amplifiers, which are of rather high importance in quantum science and technology. This can happen even if the qubit design criteria cannot be satisfied. The existing parametric amplifiers mostly are based on either JJs held at the same temperature of cryostat  or high-electron mobility transistors (HEMTs) made of III-V semiconductor heterostructures, held~typically at the operation temperature of 4 K, far above the temperature of dilution fridge. Besides the added noise as a result of higher-operation temperatures,  the usage of HEMTs  increases the complexity of circuit designs and interfacing. Existence of nonlinear quantum capacitors can offer added flexibility and convenience to this particular application. 

\section{Conclusions \& Future Work}

The graphene/boron nitride/graphene sandwich structure seems to be promising for quantum bit applications, where the nonlinearity of quantum capacitance replaces the nonlinearity of JJs. This should significantly alleviate the problem of cross-talk and altogether remove the qubit decoherence due to interference with stray magnetic fields. Using the bilayer graphene at the magic angle could be a game changer, however, the~effects of potential puddles require  careful study. It is probable that only experiments will be able to determine 
 whether such sandwich structures can envision prospects of new qubits with enhanced decoherence and dephasing properties.

As the anharmonicity stems from the quantum capacitance linked with one external linear inductor, any coupling to the circuit that modifies the effective inductance seen by the nonlinear oscillator shall modify the operating criteria. Study of an open multi-cubit system with mutual couplings and amplifiers, as well as the involved quantum circuit architecture can be a very good idea to pursue. Inductive coupling among cubits could eventually result  in useful quantum coherence with superior characteristics~\cite{Otto1,Otto2,Otto3}. Finally, the~recently developed method of higher-order operators has shown promise~\cite{CrossKerr} for analysis of nonlinear stochastic processes in quantum non-demolition measurements and quantum-limited parametric amplifiers. This could extend its possible applications to further study   nonlinear superconducting quantum circuits in quantum computing.

\vspace{6pt} 

\begin{acknowledgments}

Discussions of this work with Sungkun Hong at  the Universit\"{a}t Wien, Thomas M\"{u}ller at  the Technische Universit\"{a}t Wien, Amir Yacoby at Harvard University, John D. Teufel at the National Institute of Standards and Technology, and Cl\'{e}ment Javerzac-Galy at  the \'{E}cole Polytechnique F\'{e}d\'{e}rale de Lausanne are highly appreciated. A preliminary version of this study was presented at the Frontiers of Cricuit QED and Optomechanics, which was held on  12--14 February 2018 at  the \textit{Institute of Science and Technology}, Klosterneuburg, Austria. 

The author declares no competing or conflict of any financial and non-financial interests.

\end{acknowledgments}

\end{document}